\documentclass[sn-aps, referee]{sn-jnl}%

\usepackage{graphicx}%
\usepackage{multirow}%
\usepackage{amsmath,amssymb,amsfonts}%
\usepackage{amsthm}%
\usepackage{mathrsfs}%
\usepackage[title]{appendix}%
\usepackage{xcolor}%
\usepackage{textcomp}%
\usepackage{manyfoot}%
\usepackage{algorithm}%
\usepackage{algorithmicx}%
\usepackage{algpseudocode}%
\usepackage{listings}%
\usepackage{todonotes}
\usepackage[version=4]{mhchem}
\usepackage{siunitx}
\usepackage{hyperref}
\usepackage[subrefformat=parens]{subcaption}
\usepackage{geometry}
\usepackage{ulem}
\usepackage{threeparttable}
\usepackage{booktabs}%
\usepackage{xr}
\usepackage{physics}
\usepackage{multirow}
\usepackage{orcidlink}
\usepackage{tabularx}

\theoremstyle{thmstyleone}%

\theoremstyle{thmstyletwo}%

\theoremstyle{thmstylethree}%

\begin{document}

\title{Mechanisms and Stability of Li Dynamics in Amorphous Li-Ti-P-S-Based Mixed Ionic–Electronic Conductors: A Machine Learning Molecular Dynamics Study}

\author[1,2]{\fnm{Selva Chandrasekaran} \sur{Selvaraj}\orcidlink{0000-0002-9023-4075}}
\author[3]{\fnm{Daiwei} \sur{Wang}\orcidlink{0000-0003-2811-1071}}
\author*[3]{\fnm{Donghai} \sur{Wang}\orcidlink{0000-0001-7261-8510} \email{donghaiwang@smu.edu}}
\author*[1,2]{\fnm{Anh} \sur{T. Ngo} \email{anhngo@uic.edu}}

\affil[1]{\orgname{Department of Chemical Engineering, University of Illinois Chicago, Chicago, IL 60608, United States}}%
\affil[2]{\orgname{Materials Science Division, Argonne National Laboratory, Lemont, IL 60439, United States}}%
\affil[3]{\orgname{Department of Mechanical Engineering, Southern Methodist University, Dallas, TX 75205, USA}}%

\abstract{
	Mixed ionic-electronic conductors (MIECs) exhibit both high ionic and electronic conductivity to improve the battery performance.  In this work, we investigate the mechanism and stability of transport channels in our recently developed MIEC material, amorphous Ti-doped lithium phosphorus sulfide (LPS), using molecular dynamics (MD) simulations with a 99\% accurate machine-learning force field (MLFF) trained on \textit{ab-initio} MD data. The achieved MLFF helps efficient large-scale MD simulations on LPS with three Ti concentrations (10\%, 20\%, and 30\%) and six temperatures (25$^\mathrm{o}$C to 225$^\mathrm{o}$C) to calculate ionic conductivity, activation energy, Li-ion transport mechanism, and configurational entropy. Results show that ionic conductivities and activation energies are consistent with our recent experimental values. Moreover, Li-ion transport occurs via free-volume diffusion facilitated by the formation of disordered Li-S polyhedra. The enhanced stability of transport channels at 10\% and 20\% Ti doping, compared to 0\% and 30\%, is observed by analyzing the vibrational and configurational entropy of these disordered Li-S polyhedra. Overall, this study highlights the utility of MLFF-based large-scale MD simulations in explaining the transport mechanism and the stability of Li-ion in Ti-doped LPS electrolyte with significant computational efficiency.
}


\maketitle

\section{Introduction}

Sulfide-based solid electrolytes such as Li$_{6}$PS$_{5}$Cl, Li$_{10}$GeP$_{2}$S$_{12}$, Li$_{5.5}$PS$_{4.5}$X$_{1.5}$(x=Cl and Br), and Li$_{9.54}$Si$_{1.74}$P$_{1.44}$S$_{11.7}$Cl$_{0.35}$ demonstrate $>$10 mS/cm room temperature conductivity and high performance of all solid-state Li-ion battery (ASSLB)\cite{Li, wang, Kamaya, Adeli, Kato}. However, issues like poor cycling, low Coulombic efficiency, and rate capability under high cathode mass loading persist due to unfavorable chemistry, electrochemistry, and mechanics between the Li metal and sulfide SE \cite{Liang, Cao}. One of many approaches to overcome this aimed to develop  mixed ionic-electronic conductive (MIEC) solid electrolytes\cite{So}. 

A MIEC in crystalline or amorphous state represents a promising class of materials for next-generation energy storage devices, including lithium-ion and lithium-sulfur (Li-S) batteries for more than one decade\cite{Cao, Abdulkadir, Lin}. These materials exhibit both ionic and electronic conductivity, enabling them to play a dual role in facilitating lithium-ion transport while simultaneously providing pathways for electronic charge transfer. Unlike traditional solid electrolytes, which are purely ionic conductors, MIECs improve overall cell efficiency by addressing critical challenges such as sluggish redox kinetics, insulating discharge products, and interfacial instabilities\cite{Ma}. Especially, the amorphous structure further enhances their ionic conductivity by reducing crystallographic barriers to ion migration, making them highly attractive for solid-state applications\cite{Patrick}.

A notable example and our recent development of MIEC is amorphous titanium-doped lithium phosphorus sulfide (Li-Ti-P-S), which exhibits excellent ionic conductivity (10$^{-4}$–10$^{-3}$ S cm$^{-1}$ at 25 $^o$C)  due to its disordered sulfide-based structure and moderate electronic conductivity \cite{donghai}. The dual functionality of this MIEC solid electrolyte makes it particularly suitable for solid-state lithium-sulfur batteries, where it helps to high discharge capacity ($>$1450 mAh/g) and extend cycle life ($>$1000 cycles)\cite{donghai}.  At the same time, mechanism and stability of Li dynamic channels are unclear from our experimental work  \cite{donghai} when Li-P-S based ionic conductor become Li-Ti-P-S based MIEC solid electrolyte. 

To address this challenge, we proposed a large-scale dynamics study based on a machine-learning force field (MLFF) to investigate the mechanism and stability of lithium dynamic channels in the Li-Ti-P-S electrolyte\cite{selva_ltc}. Large-scale simulations are essential for this study as they capture the complex Li-ion dynamics in the amorphous state, which cannot be adequately represented by small-scale approaches. While small-scale dynamics studies using density functional theory (DFT) are well-established for predicting structural properties, electrochemical behavior, and Li-ion transport barriers, they are typically limited to systems with a few hundred atoms\cite{selva, selva1}. Extending such simulations to the large-scale systems required for modeling amorphous Li-dynamics would demand prohibitively high computational resources. Therefore, DFT is not a practical choice for this purpose. Instead, MLFF-based approaches provide a computationally efficient alternative to bridge this gap.

\begin{figure}
	\centering
	\includegraphics[width=\textwidth]{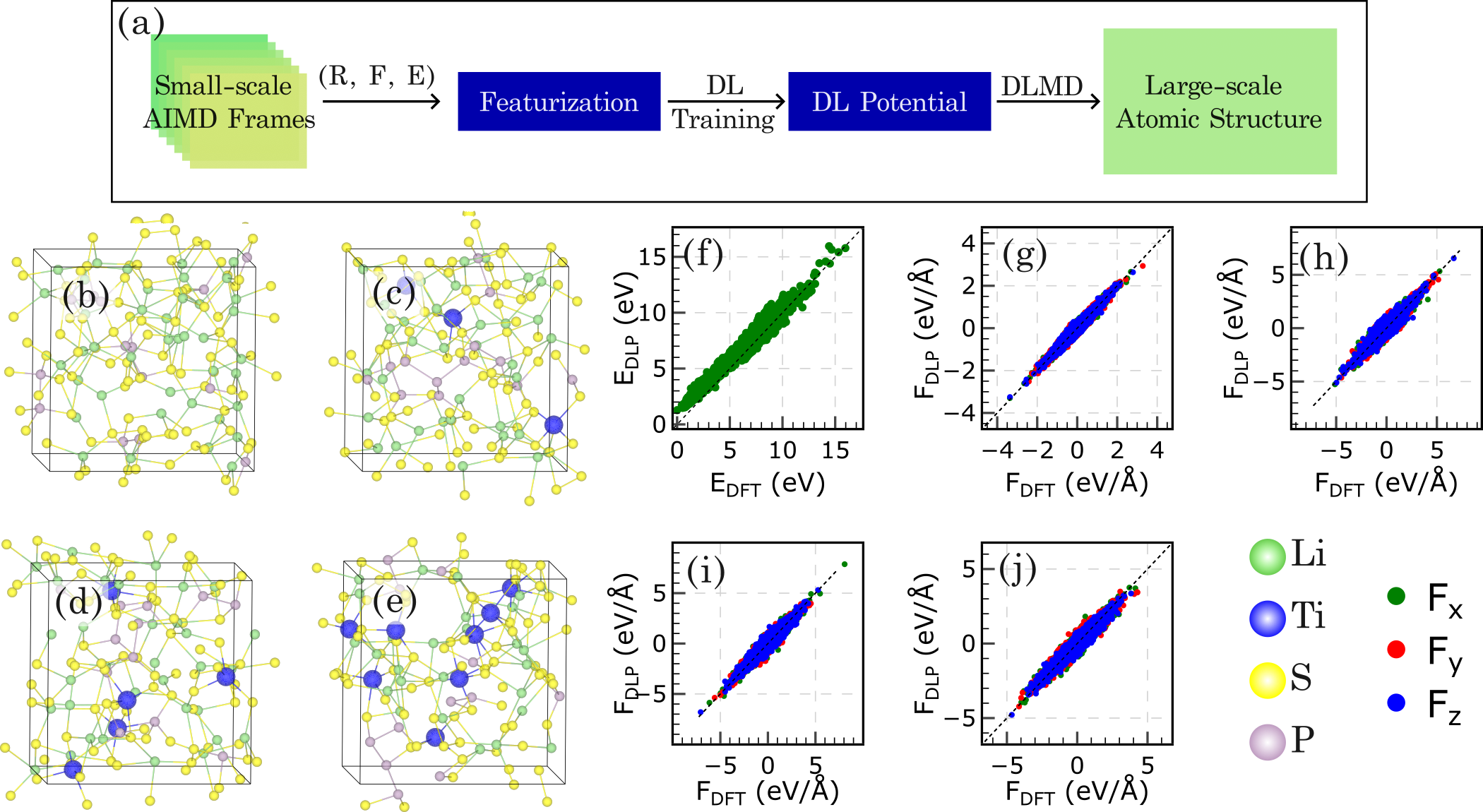}
	\vspace{1mm}
	\caption{(a) Schematic diagram of the protocol of DLMD development (see \autoref{schematic}), where R, F, E, and DL represent atomic coordinates, forces, energies, and deep learning, respectively. (b-e) Ball and stick models of amorphous LPS(b), LPS:Ti$_{10\%}$(c), LPS:Ti$_{20\%}$(d), and LPS:Ti$_{30\%}$(e), respectively. (f-j) Accuracy of trained force field model with DFT data versus predicted data such as energy(f) and forces along F$_x$, F$_y$, and F$_z$ of Li(g), P(h), S(i), and Ti(j). }
	\label{FIG1}
\end{figure}

Our approach in this work involves the integration of DFT-based $ab-initio$ molecular dynamics (AIMD), machine learning force filed based on deep neural networks, and classical molecular dynamics to investigate the structural and Li-ion dynamics in Li-Ti-P-S electrolyte. This integrated approach is collectively referred to as deep learning molecular dynamics (DLMD) simulations and its force field is called deep learning potential (DLP) \cite{Peng_2023, Zhang}. With this, we organize the manuscript as follows:  \autoref{methodology} describes detailed simulation techniques, as illustrated in \autoref{FIG1}.  \autoref{results} is composed of results and discussions of mean square displacement (MSD), diffusion coefficient, ionic conductivity, activation energy, transport mechanism, and stability of Li-ion ions. Finally, we conclude our results and discussions in \autoref{conclusion}.

\section{Simulation Methodologies}{\label{methodology}}
\subsection{Structure Modeling}
The preparation of simulation cells for amorphous solid electrolytes was carried out to maintain compositions close to approximately 100 atoms per cell. Four different material systems were considered: (1) 75\% of \(\text{Li}_2\text{S}\) and 25\% of \(\text{P}_2\text{S}_5\) (LPS:Ti$_{00\%}$), (2) 75\% of \(\text{Li}_2\text{S}\), 25\% of \(\text{P}_2\text{S}_5\), and 10\% of \(\text{TiS}_2\) (LPS:Ti$_{10\%}$), (3) 75\% of \(\text{Li}_2\text{S}\), 25\% of \(\text{P}_2\text{S}_5\), and 20\% of \(\text{TiS}_2\) (LPS:Ti$_{20\%}$), and (4) 75\% of \(\text{Li}_2\text{S}\), 25\% of \(\text{P}_2\text{S}_5\), and 30\% of \(\text{TiS}_2\) (LPS:Ti$_{30\%}$). The composition of each system was calculated based on the molar ratios of the constituent materials.

For the first case, the atomic contributions of Li, P, and S were determined based on their respective molecular formulas. To achieve a target composition with a total atom count close to 100, a ratio of 3 formula units of \(\text{Li}_2\text{S}\) to 1 formula unit of \(\text{P}_2\text{S}_5\) was selected, reflecting the 3:1 molar ratio (75\%:25\%). This combination contributes a total of \(3 \times 3 + 1 \times 7 = 16\) atoms per set. To approximate 100 atoms, the ratio was scaled by calculating the total number of formula units as \(\frac{100}{16} \approx 6\). Using this scaling, a simulation cell containing 96 atoms (\(6 \times 16 = 96\)) was constructed for the LPS:Ti$_{00\%}$ system. With these 96 atoms, an initial cubic structure was generated using Atomsk software~\cite{atomsk}, where the volume was approximated based on the given molar ratio, and an interatomic distance of $2\text{\AA}$ was maintained.

Similarly, the compositions and simulation cells of LPS:Ti$_{10\%}$, LPS:Ti$_{20\%}$, and LPS:Ti$_{30\%}$ were constructed. For these systems, \(\text{TiS}_2\) was introduced in proportions of 10\%, 20\%, and 30\%, respectively. The molar ratios of the constituent materials were adjusted accordingly to incorporate Ti into the simulation cells while ensuring the total atom count remained close to 100. This systematic approach ensured that the simulation cells maintained the specified molar ratios for all four compositions.

\begin{table}[h]
	\centering
	\caption{ The calculated loss function parameter expressed by mean absolute error (MAE) and  root mean square error (RMSE) of deep learning potential model}
	\label{table1}
	\begin{tabularx}{\linewidth}{XcXcXc}
		\toprule
		\textbf{Metric} 	& \textbf{Value} 				    & \textbf{Unit} &\\
		\midrule
		Energy MAE    	   & $6.723 \times 10^{-4}$  & eV/atom 	&	\\
		Energy RMSE 	  & $8.389 \times 10^{-4}$ & eV/atom 	&   \\
		\midrule
		Force MAE   	    & $5.959 \times 10^{-3}$ & eV/\AA 	&	\\
		Force RMSE 		   & $7.827 \times 10^{-3}$ & eV/\AA  	&	\\
		\bottomrule
	\end{tabularx}
\end{table}

\subsection{Data preparation}\label{schematic}
DFT-based AIMD calculations were employed to optimize the atomic positions and the structures of the initially designed LPS:Ti$_{00\%, 10\%, 20\%, 30\%}$ systems using the Vienna $Ab~initio$ Simulation Package (VASP)~\cite{vasp1, vasp2}. The projector augmented wave (PAW) formalism, which describes the valence electrons of Li, P, S, and Ti atoms using plane wave-based wave functions, was employed~\cite{paw}. Structure optimization, involving the minimization of the ground-state energy, utilized the generalized gradient approximation method of Perdew and Wang~\cite{gga, gga1}. A kinetic energy cutoff of 600~eV was set to enhance calculation accuracy. Ionic and electronic optimizations were alternately performed until the forces on each ion were less than ±10~meV/Å.  The ball and stick models of the optimized  LPS:Ti$_{00\%, 10\%, 20\%, 30\%}$ structures were given in \autoref{FIG1} (b-e), respectively.

For the AIMD simulations, a time step of 1~fs was used to integrate the equations of motion, with an energy convergence criterion of 10$^{-6}$~eV. Additionally, an NVT ensemble was employed. The structures were heated to 1000~K at a rate of 70~K/ps, held at 1000~K for 3~ps, and then cooled at the same rate to the target temperature. Finally, simulations were performed at 500~K, 600~K, 700~K, 800~K, and 900~K for at least 50~ps to generate data for training the DLP.

\subsection{DLMD simulation}\label{schematic}
The DLP was developed using the deep neural network method implemented in DeePMD-kit (v2.2.7) \cite{zeng2023deepmd}. The deep neural network algorithm in DeePMD is designed using the TensorFlow Python library  \cite{tensorflow} with three embedding neural network layers, each containing 32, 64, and 128 neurons. The deep potential-smooth edition (DeepPot-SE), which is an end-to-end machine learning-based potential energy surface (PES) model, was employed with a cutoff radius of 7 Å to include more neighbour atoms in the featurization process. Indeed, it efficiently represents the PES of a wide variety of systems with the accuracy of AIMD calculations \cite{DeepPot-SE}.

The training process consisted of \(2 \times 10^4\) steps, utilizing mean squared error for both energies and forces as the loss function at each training step. The optimization was facilitated by the Adam optimizer, initiating with a learning rate of \(10^{-3}\) and concluding at \(10^{-8}\), with a decay parameter set to 5000. With these specified parameters, a dataset comprising more than 2 million training samples and 20000 test samples from diverse trajectories was employed to construct the DLP model. The calculated loss function parameters such as mean absolute error and root mean square error during validation of the developed DLP model are tabulated in Table-1. Accuracy of the predicted data of force and energy is more than 99 \% and data points of predicted versus trained are shown in \autoref{FIG1}f to \autoref{FIG1}j. 

\begin{figure}[htbp]
	\centering
	\includegraphics[width=\textwidth]{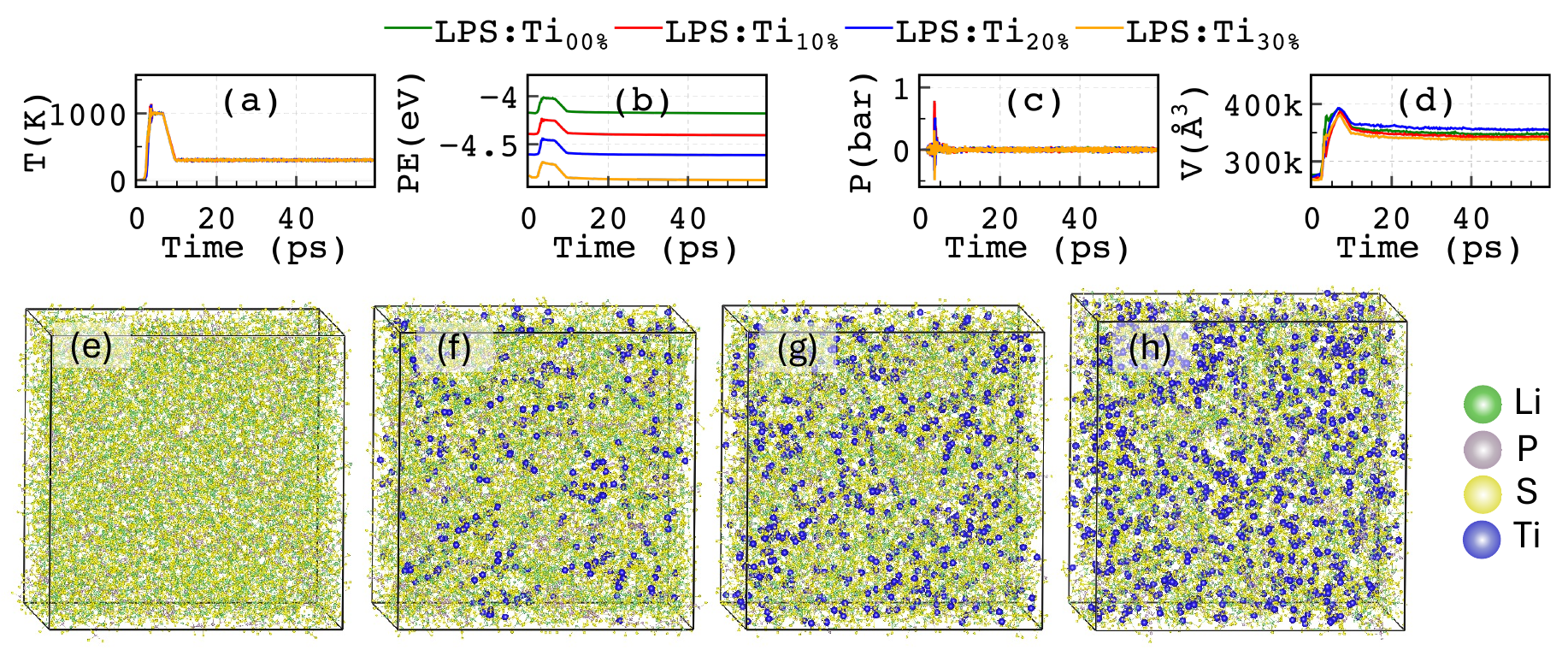}
	\vspace{1mm}
	\caption{Temperature ($T$), potential energy ($PE$), pressure ($P$), and volume ($V$) profiles from NPT simulations (panels \textbf{a--d}) and ball-and-stick models of their optimized structures (\textbf{e--h}) of LPS:Ti$_{00\%, 10\%, 20\%, 30\%}$.}
	\label{FIG2}
\end{figure}	

\section{Result and Discussion}{\label{results}}

\subsection{Structure optimization}
Using the developed DLP, DLMD simulations were performed in the isothermal-isobaric (NPT) ensemble to control the temperature and pressure of the system and optimize a large structure consisting of 12,000 atoms, designed from the output of AIMD. The simulation process began with an initial phase at a low temperature of 1~K for 2~ps to stabilize the initial configuration. Subsequently, the system was gradually heated from 1~K to 1000~K over another 2~ps, allowing a smooth transition to a higher thermal state. After reaching 1000~K, the system was equilibrated at this elevated temperature for 3~ps to ensure thermal and pressure stability. A controlled cooling phase was then implemented, reducing the temperature from 1000~K to 300~K over 3~ps. Finally, the system was equilibrated at 300~K for 50~ps to simulate standard room-temperature conditions. The temperature, potential energy, pressure, and volume profiles with respect to time for LPS:Ti$_{00\%, 10\%, 20\%, 30\%}$ systems are plotted in \autoref{FIG2}~(a-d) and their optimized structures are shown in \autoref{FIG2}(e-h).  All their simulations were done using the LAMMPS simulation package \cite{lammps, lammps1, lammps2} coupled with the DeepMD plugin \cite{WANG2018178}. 

With the optimized large structure, DLMD simulations were conducted in the NVT ensemble to investigate the Li-dynamic properties. Simulations were performed at eight different temperatures such as 300, 320, 340, 360, 380, 400, 450, and 500~K (25$^\mathrm{o}$C to 225$^\mathrm{o}$C), controlled using the Nosé-Hoover thermostat. The temperature damping parameter was set to 100~fs, and a uniform integration time step of 1~fs was employed for all simulations, extending over a total duration of 3.0~ns. The small and large cells from the AIMD and DLMD simulations were visualized using VESTA software\cite{vesta}, as shown in \autoref{FIG1}~(b-e) and \autoref{FIG2}~(e-h), respectively. Additionally, the trajectories of the AIMD and DLMD simulations were analyzed and visualized using OVITO software.\cite{Ovito}

\begin{figure}[htbp]
	\centering
	\includegraphics[width=13cm]{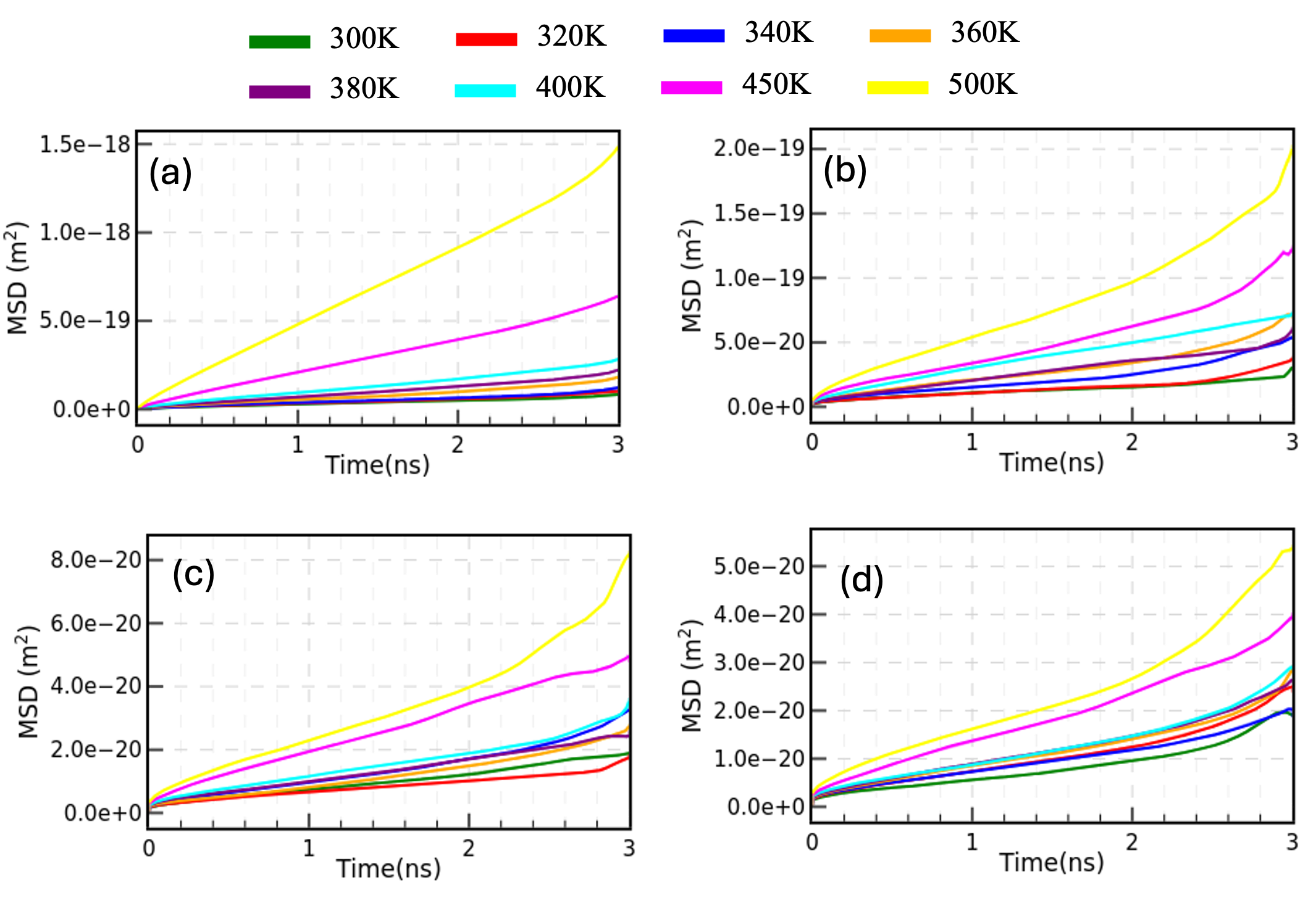}
	\vspace{0.5mm}
	\caption{MSD  of LPS:Ti$_{00\%}$(a), LPS:Ti$_{10\%}$(b), LPS:Ti$_{20\%}$(c), and LPS:Ti$_{30\%}$(d).}
	\label{FIG3}
\end{figure}	

\subsection{Mean Square Displacement Analysis}\label{msd}
The mean square displacement (MSD) as a function of time, calculated over a simulation duration of 3.0 ns from the LAMMPS trajectory, provides critical insights into the dynamic behavior of particles within the system. The MSD is defined as the average squared displacement of particles from their initial positions and is given by \autoref{msd}:

\begin{equation}\label{msd}
	\text{MSD}(t) = \frac{1}{N} \sum_{i=1}^N \langle |\mathbf{r}_i(t) - \mathbf{r}_i(0)|^2 \rangle
\end{equation}
\noindent
where \(N\) is the total number of particles, \(\mathbf{r}_i(t)\) represents the position of particle \(i\) at time \(t\), and \(\langle \cdot \rangle\) denotes ensemble averaging. The calculated MSD values for temperatures and different Ti concentrations are plotted in \autoref{FIG3}. 

To analyze the diffusion regime, we calculated the logarithmic slope \( \beta = \frac{d(\log \text{MSD})}{d(\log t)} \). The \( \beta \)-value analysis provides important insights into the Li-ion dynamics. When \( \beta < 1 \), it indicates restricted or hindered motion of Li-ions due to structural constraints, such as defects, grain boundaries, or trapping effects. A \( \beta \)-value of 1 reflects normal diffusion, where the motion corresponds to random Brownian motion consistent with Fickian diffusion. For \( 1 < \beta < 2 \), the motion becomes super-diffusive, suggesting enhanced ion motion due to cooperative effects or highly conductive pathways. At very short timescales, a \( \beta \)-value of 2 corresponds to ballistic motion, where particle motion is dominated by inertia before interactions with the surrounding environment. 

We chose the time range from 0.2 to 2.6~ns for the analysis, which provided \( \beta \)-values between 0.99 and 1.1, indicating a transition between normal diffusion and slightly super-diffusive behavior. The observed \( \beta \)-values in this study, which are close to 1, suggest predominantly normal diffusion behavior for Li-ions in the system, with slight deviations toward super-diffusion under certain conditions. These results provide critical insights into the material's ability to facilitate Li-ion transport under varying conditions.

\begin{figure}[htbp]
	\centering
	\includegraphics[width=\textwidth]{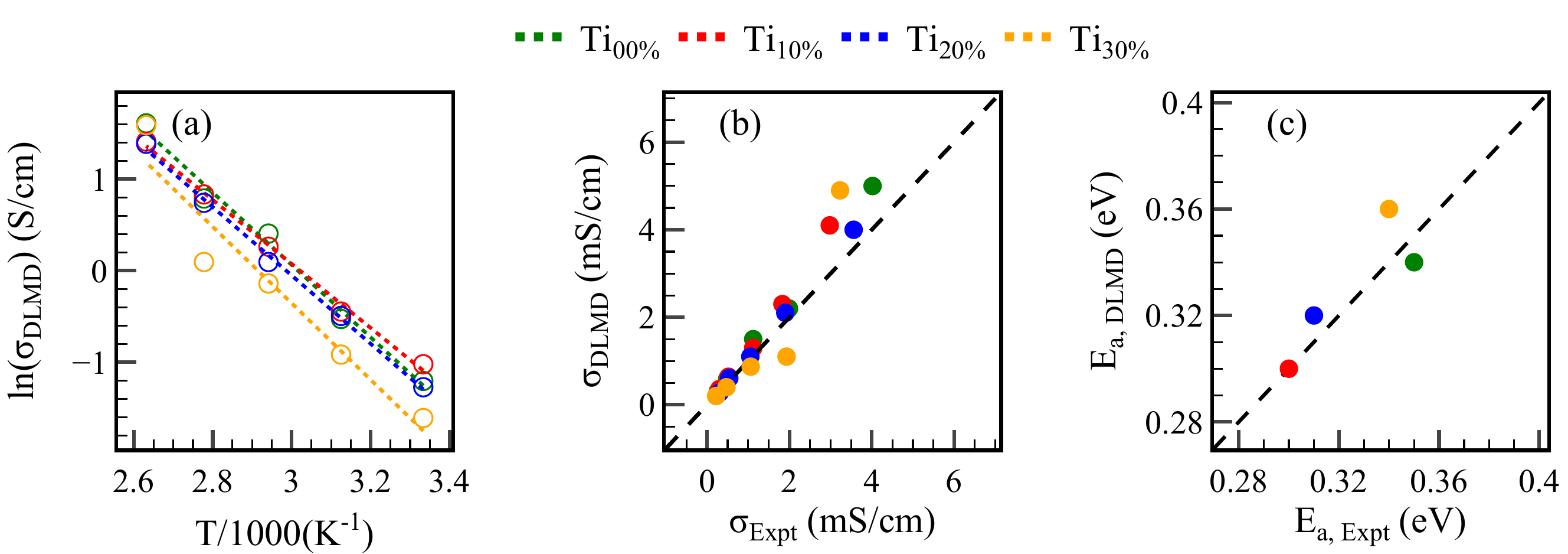}
	\vspace{1mm}
	\caption{(a) Calculated ionic conductivities of LPS:Ti at different temperatures: 300, 320, 340, 360, and 380~K (b) Calculated versus experimental values of ionic conductivity.  (c) Calculated versus experimental values of activation energy of  LPS:Ti Compounds.}
	\label{FIG4}
\end{figure}

\subsection{Ionic Conductivity and Activation Energy}
From the MSD values corresponding to the diffusive regime, the diffusion coefficient (\(D\)) of Li-ions in LPS:Ti was calculated using ~\autoref{diff_co}:

\begin{equation}\label{diff_co}
	D = \frac{{\rm MSD}(\Delta t)}{2d\Delta t}
\end{equation}
\noindent
where \(d\) is a dimensionless quantity equal to 3 for three-dimensional transport. 

The ionic conductivity, denoted as $\sigma$, can be determined from the diffusion coefficient $D$ (see \autoref{diff_co}) using the Nernst--Einstein equation \cite{sigma}:
\begin{equation}
	\sigma = \frac{n e^2 Z^2}{k_{\mathrm{B}} T} D
\end{equation}
\noindent
where $n$ represents the number density of Li ions (\si{\per\cubic\meter}), $e = \SI{1.602e-19}{\coulomb}$ is the elementary charge, $Z = +1$ is the valence of Li, $k_{\mathrm{B}} = \SI{1.381e-23}{\joule\per\kelvin}$ is the Boltzmann constant, and $T$ is the absolute temperature (\si{\kelvin}). 

The calculated ionic conductivity values as a function of Ti concentration and temperature are shown in \autoref{FIG4}a and align well with our recent  experimental data \cite{donghai}, as demonstrated in \autoref{FIG4}b. To estimate the activation energy $E_{\mathrm{a}}$, the temperature dependence of $\sigma$ is fitted to the Arrhenius relationship:
\begin{equation}\label{sigma}
	\sigma = \sigma_0 e^{-E_{\mathrm{a}}/(k_{\mathrm{B}}T)}
\end{equation}
\noindent
where $\sigma_0$ is the pre-exponential factor (determined from the $y$-intercept of $\ln(\sigma T)$ versus $1/T$ plot) and $E_{\mathrm{a}} = -m k_{\mathrm{B}}$ is derived from the slope $m$ of $\ln(\sigma T)$ versus $1/T$, with $k_{\mathrm{B}} = \SI{8.617e-5}{\electronvolt\per\kelvin}$ when expressing $E_{\mathrm{a}}$ in \si{\electronvolt}.

The calculated activation energies, shown in ~\autoref{FIG4}c, closely match experimental values. Interestingly, 10\% Ti in LPS produces an activation energy of 0.3~eV, which is the lowest among all compositions. The 20\% Ti composition shows an activation energy of 0.32~eV, also close to that of the 10\% Ti case. However, compositions with 0\% Ti and 30\% Ti exhibit higher activation barriers compared to the 10\% and 20\% Ti cases. This indicates that adding more than 20\% Ti or excluding Ti entirely is not favorable for achieving low activation energy and, thus, efficient Li-ion conduction.

\begin{figure}[htbp]
	\centering
	\includegraphics[width=14cm]{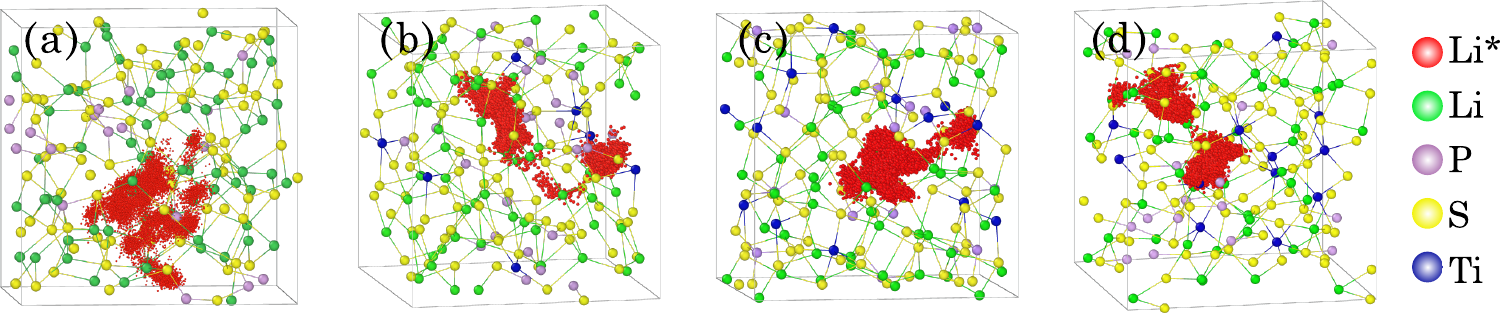}
	\includegraphics[width=14cm]{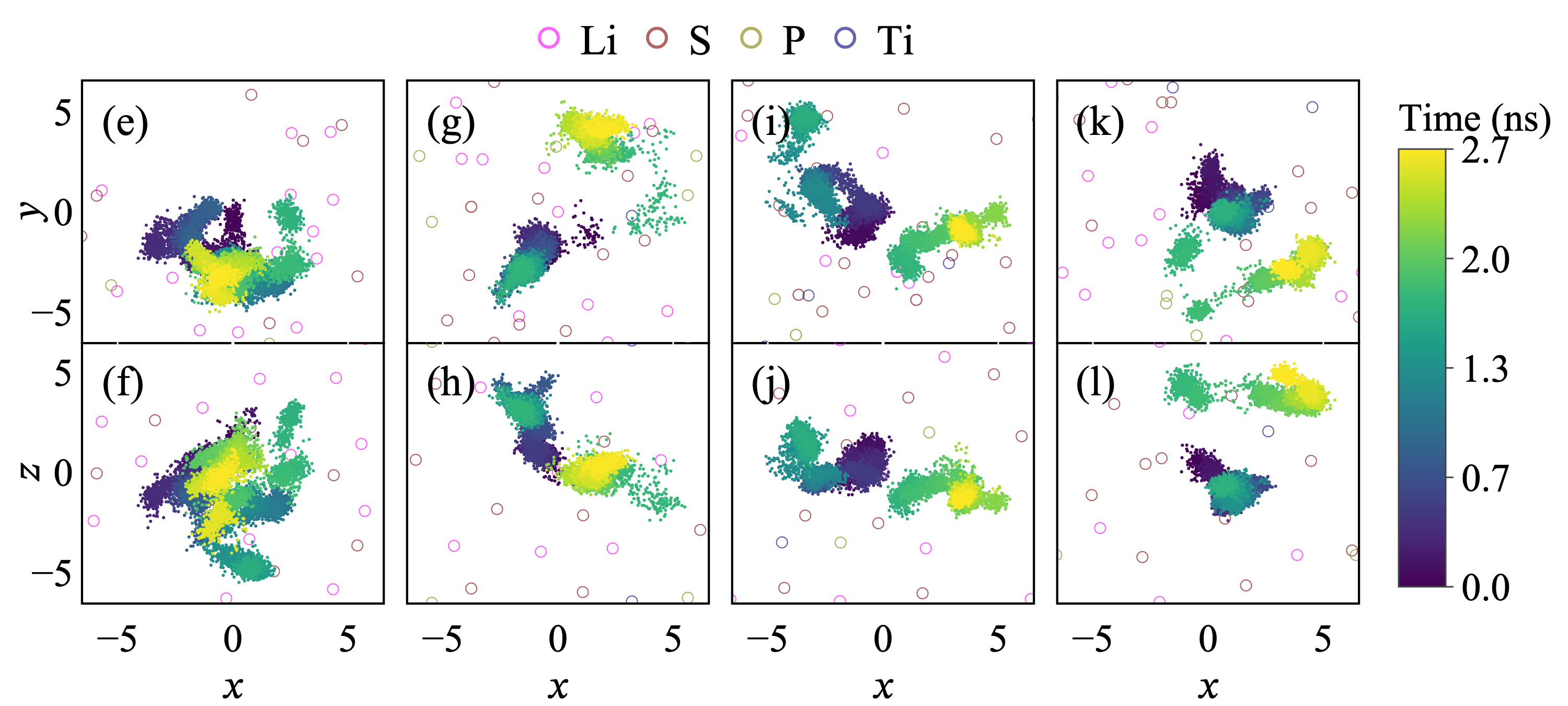}
	\vspace{1mm}
	\caption{ Ball and stick model for moving Li-ion(Red, Li*) with respect to time step  of LPS:Ti$_{00\%}$(a), LPS:Ti$_{10\%}$(b), LPS:Ti$_{20\%}$(c), and LPS:Ti$_{30\%}$(d). Here, the cubic boundaries of the ball and stick models are psudo-boundaries that are only to show the movement of one particular Li-ion in a large cell. The Li-ion movement in 2D plane ($xy$ and $xz$ in \AA) with respect to time steps is plotted in (e-l). }
	\label{FIG5}
\end{figure}	

\subsection{Li-ion Transport Mechanism}
Li-ion transport in the LPS solid electrolyte occurs via a free-volume diffusion mechanism. This mechanism implies that Li-ions migrate through the disordered or amorphous structure by hopping into neighboring regions of unoccupied space, known as free-volume. In the amorphous structure of LPS, the absence of long-range crystalline order creates regions of varying density, with some areas containing excess space or voids. These voids serve as pathways for particle diffusion. \autoref{FIG5} illustrates 3D and 2D models of Li-ion migration through voids within a specific region over the simulation time. \autoref{FIG5} confirms that the migration paths of Li-ions in LPS with 10\% and 20\% Ti doping cover more area in 2D or volume in 3D compared to 0\% and 30\% Ti doping. This indicates that Li-ion movement in LPS with 10\% and 20\% Ti doping is significantly easier than in the other two cases.

Furthermore, the free-volume diffusion process in LPS involves the formation of disordered Li-S$_n$ ($n$ = 1 to 6) polyhedra within the electrolyte matrix. The disordered nature of these polyhedra reduces structural constraints which results in variations in the migration speed of Li-ions and affects the stability of Li-ion migration channels. Therefore, a quantitative analysis of Li-S polyhedra in Ti-doped LPS is crucial for understanding the stability and effectiveness of Li-ion transport channels.

\subsection{Stability of transport channel}
\subsubsection{Li-S Polyhedra Coordination Analysis}\label{Li-S Polyhedra Coordination Analysis} 
The coordination probability of the Li-S pair was calculated by analyzing the trajectories obtained from DLMD simulations. For each frame of the trajectory, the positions of lithium (Li) and sulfur (S) atoms were extracted to determine the local coordination number (\(n_c\)) of Li atoms. A range of cutoff distances, \( r_{\mathrm{cut}} \) (2.5 to 3.1 \AA), was applied to identify neighboring S atoms around each Li atom.

Using the \( k \)-nearest neighbor ($k$-NN) method, the number of neighboring S atoms (\(N_S\)) within \( r_{\mathrm{cut}} \) for each Li atom was determined. In this method, the number of neighbors \( n_c \) was computed as:

\begin{equation}\label{nc}
n_c = \sum_{i=1}^{N_S} \mathcal{I}\left(d(\mathbf{x}, \mathbf{x}_i) \leq r_{\mathrm{cut}}\right),
\end{equation}
\noindent
where \( N_S \) represents the total number of S atoms within the cutoff radius from a Li atom, \( d(\mathbf{x}, \mathbf{x}_i) \) is the Euclidean distance between the position \( \mathbf{x} \) of a Li atom and the position \( \mathbf{x}_i \) of an S atom, and \( \mathcal{I}\left(d(\mathbf{x}, \mathbf{x}_i) \leq r_{\mathrm{cut}}\right) \) in \autoref{nc} is an indicator function defined as in \autoref{nc value}:

\begin{equation}\label{nc value}
\mathcal{I}\left(d(\mathbf{x}, \mathbf{x}_i) \leq r_{\mathrm{cut}}\right) =
\begin{cases}
	1, & \text{if } d(\mathbf{x}, \mathbf{x}_i) \leq r_{\mathrm{cut}}, \\
	0, & \text{otherwise.}
\end{cases}
\end{equation}

This approach ensures that only those S atoms within the specified cutoff distance \( r_{\mathrm{cut}} \) contribute to the coordination number of each Li atom. For every trajectory frame, the coordination number distribution was quantified by counting the frequency of neighboring S atoms for all Li atoms and normalizing this count by the total number of Li atoms in the system, \( N_{\mathrm{Li}} \). This normalization yielded the coordination probability:

\begin{equation}\label{probability}
P(n_c) = \frac{N(n_c)}{N_{\mathrm{Li}}},
\end{equation}
\noindent
where \( N(n_c) \) is the number of Li atoms with a specific coordination number, ranging from 0 to 6.

Furthermore, the $k$-NN algorithm facilitated frame-by-frame analysis of Li-S coordination across the entire trajectory. This approach allowed for the dynamic variation in coordination environments to be captured and analyzed over time. The results provided insights into the structural stability and ionic transport properties of the material, which are crucial for understanding the behavior of Li atoms in solid-state electrolyte systems.

\begin{figure}[htbp]
	\centering
	\includegraphics[width=12cm]{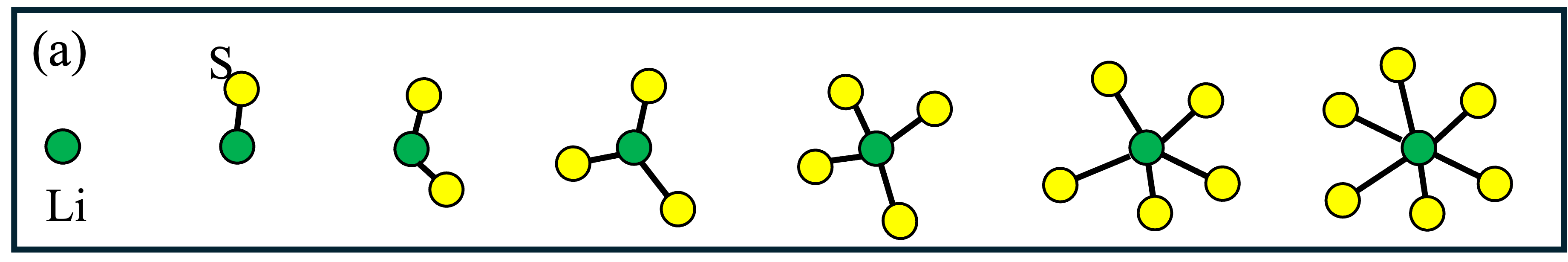}
	\includegraphics[width=\textwidth]{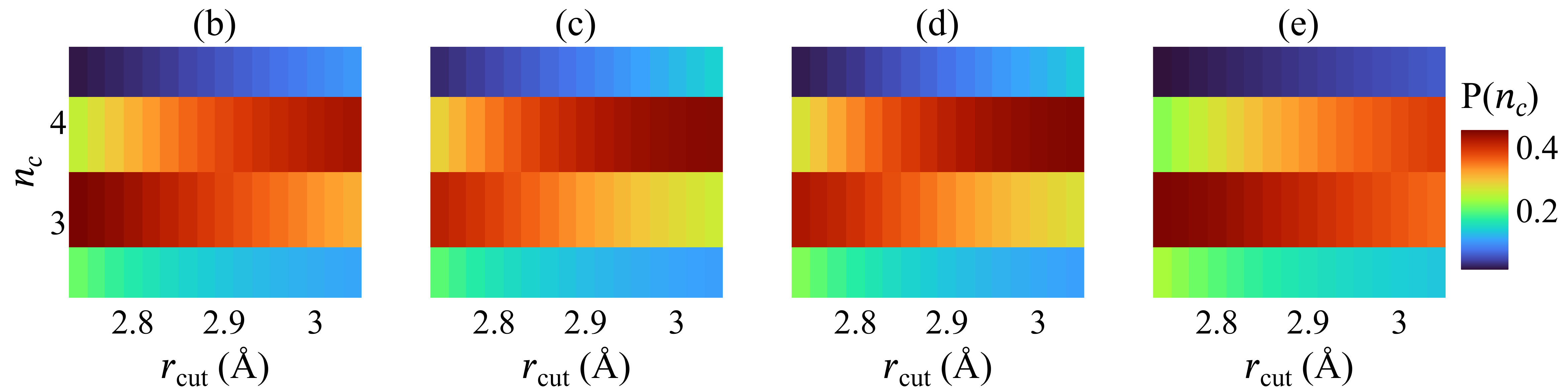}
	\vspace{1mm}
	\caption{(a) 2D ball-and-stick model of Li-S polyhedra with coordination numbers of 0, 1, 2, 3, 4, 5, and 6. Coordination numbers (\(n_c\) as CN) as a function of cutoff radius for LPS:Ti$_{00\%}$ (a), LPS:Ti$_{10\%}$ (b), LPS:Ti$_{20\%}$ (c), and LPS:Ti$_{30\%}$ (d). The coordination probabilities are given in the color maps on the y-axis.}
	\label{FIG6}
\end{figure}	

 \autoref{FIG6} illustrates heatmaps of coordination probabilities for different Ti doping levels as a function of the cutoff distance, \( r_{\mathrm{cut}} \). The analysis revealed several important trends. For Ti doping levels of 10\% and 20\%, more than 50\% of Li atoms were found in 4-coordinated environments with S atoms. This indicates a stable and efficient Li-S polyhedral network favorable for Li-ion transport. In contrast, for Ti doping levels of 0\% and 30\%, more than 50\% of Li atoms were 3-coordinated with S atoms. This suggests less stable structures, which may hinder Li-ion migration.

\subsubsection{Configurational Entropy}
In thermodynamics, the total entropy of a system comprises contributions from multiple sources as defined in \autoref{entropy}:

\begin{equation}\label{entropy}
S_{\text{total}} = S_{\text{trans}} + S_{\text{elec}} + S_{\text{vib}} + S_{\text{config}},
\end{equation}
\noindent
where \( S_{\text{trans}} \) is the translational entropy, associated with the motion of ions in space; \( S_{\text{vib}} \) is the vibrational entropy, arising from atomic vibrations; \( S_{\text{elec}} \) is the electronic entropy, accounting for electronic excitations; and \( S_{\text{config}} \) is the configurational entropy, which in this study explicitly refers to the degree of disorder in the Li–S polyhedra within the amorphous Li-Ti-P-S matrix.

The vibrational and configurational entropy in \autoref{entropy} was derived from the probabilities of different coordination environments using the Boltzmann entropy formula:

\begin{equation}\label{entropy_config}
	S_{\mathrm{config}} = -k_{\mathrm{B}} \sum_{n_c} P(n_c) \ln P(n_c),
\end{equation}

\begin{equation}\label{entropy_elect}
S_{\mathrm{vib}} = k_B \sum_{i} g(\omega_i) \left[ \frac{\beta \hbar \omega_i}{e^{\beta \hbar \omega_i} - 1} - \ln\left(1 - e^{-\beta \hbar \omega_i}\right) \right]
\end{equation}
\noindent
where \( k_{\mathrm{B}} \) is the Boltzmann constant, \( P(n_c) \) represents the probability of each coordination number, \( n_c \) ranges from 1 to 6, \( \hbar \) is reduced Planck's constant, \( \beta = \frac{1}{k_B T} \) is inverse thermal energy, \( T \) is temperature, \( \omega_i \) is vibrational angular frequency (\( 2 \pi \times \text{frequency} \)), and \( g(\omega_i) \) is vibrational density of states (VDOS). To ensure numerical stability, a small value (\(1 \times 10^{-12}\)) was added to \( P(n_c) \) in \autoref{entropy_config} if it was zero.

The calculated vibrational entropy value is in the range of 1 to \(3mk_{\mathrm{B}} \), whereas configurational entropy is in the order of  \(1k_{\mathrm{B}} \).  This reveals that the configurational entropy becomes particularly significant in systems where ionic transport is strongly influenced by the structural arrangement and disorder of these polyhedra, whereas the contributions from vibrational, translational, and electronic entropies are negligibly small.  The Gibbs free energy \( \Delta G \) of a system can be written as \(\Delta G = H - T S_{\text{total}}\), where \( H \) is the enthalpy, \( T \) is the temperature, and \( S_{\text{total}} \) is the total entropy. Focusing on the configurational entropy contribution, the Gibbs free energy mostly depends on \( S_{\text{config}} \). When \( S_{\text{config}} \) increases, \( \Delta G_{\text{config}} \) decreases, which indicates a more thermodynamically favorable state for ionic transport.

\begin{figure}[htbp]
	\centering
	\includegraphics[width=0.5\textwidth]{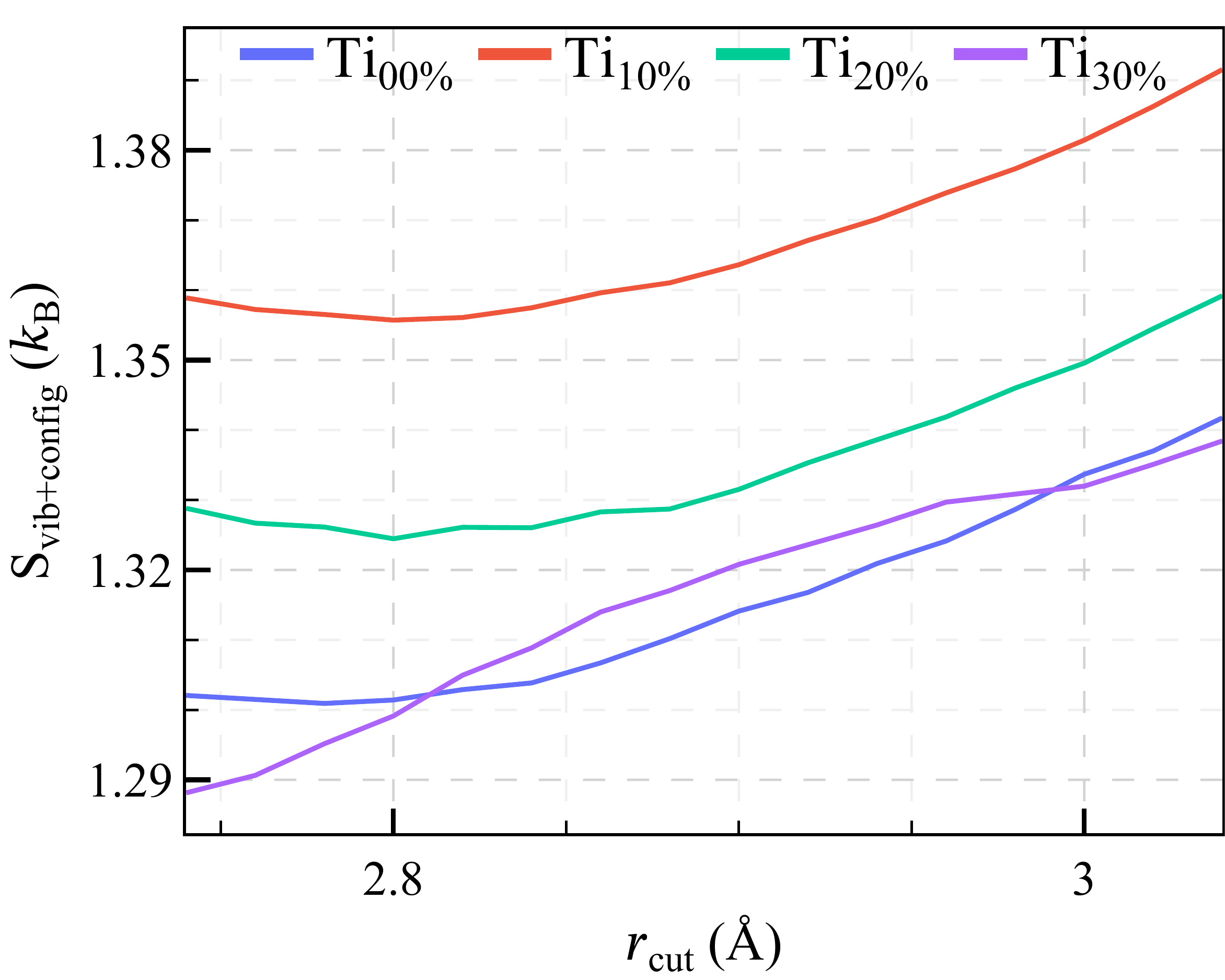}
	\vspace{1mm}
	\caption{The calculated vibrational and configurational entropy in units of \( k_{\mathrm{B}} \) is plotted as a function of cutoff radius for LPS:Ti$_{00\%}$, LPS:Ti$_{10\%}$, LPS:Ti$_{20\%}$, and LPS:Ti$_{30\%}$.}
	\label{FIG7}
\end{figure}

The configurational entropy as a function of \( r_{\mathrm{cut}} \) is plotted in  \autoref{FIG7} which shows Ti doping at 10\% exhibited the highest entropy, followed by 20\%. The increased entropy in these cases correlates with higher structural disorder with more stable four-coordinated Li atoms (\autoref{Li-S Polyhedra Coordination Analysis} and \autoref{FIG6}), which enhances the free-volume diffusion mechanism. At the same time, higher configurational entropy corresponds to a decrease in Gibbs free energy, making the system more thermodynamically favorable for Li-ion diffusion. Specifically,  10\% and 20\% of Ti doping reduce the Gibbs free energy and they enhance the driving force for ionic transport. This demonstrates the importance of Ti doping concentration in tuning the Li-ion transport mechanism.

On the other hand, Ti doping levels of 0\% and 30\% exhibited significantly lower configurational entropy, which leads to comparatively higher Gibbs free energy and material instability. Therefore, Ti doping at 10\% and 20\% optimizes both the coordination environment and configurational entropy, leading to enhanced ionic conductivity and reduced activation energy barriers, ultimately improving the overall performance of the solid-state electrolyte.

\section{Conclusion} \label{conclusion}

This study demonstrates the utility of MIEC solid electrolytes in enhancing battery performance by improving ionic and electronic conductivity. Using MD simulations with a 99\% accurate MLFF trained on \textit{ab-initio} AIMD data, we investigated the transport mechanism and stability of Ti-doped LPS across three doping concentrations (10\%, 20\%, and 30\%) and different temperatures (25$^\mathrm{o}$C to 225$^\mathrm{o}$C). The MLFF-enabled large-scale MD simulations provided efficient and accurate calculations of ionic conductivity, activation energy, Li-ion transport mechanism, and configurational entropy. The results revealed that the ionic conductivities and activation energies were consistent with our experimental values, validating the reliability of the MLFF approach. Li-ion transport in the Ti-doped LPS electrolyte was found to occur via a free-volume diffusion mechanism, facilitated by the formation of disordered Li-S polyhedra. 

The configurational entropy analysis of these disordered Li-S polyhedra highlighted enhanced transport stability at 10\% and 20\% of Ti doping compared to 0\% and 30\%. This suggests that optimal doping levels (10\% and 20\%) improve the structural disorder necessary for effective ionic transport while maintaining thermodynamic stability. Overall, this work underscores the capability of MLFF-based large-scale MD simulations in capturing the complex transport dynamics and stability of Li-ion in Ti-doped LPS electrolytes with significant computational efficiency. These findings provide valuable insights into the design of high-performance MIECs and solid electrolytes for next-generation battery technologies.

\backmatter

\section*{Acknowledgements}
This work was supported by the Assistant Secretary for Energy Efficiency and Renewable Energy, Office of Vehicle Technologies of the US Department of Energy, through the Battery Materials Research (BMR) program. We gratefully acknowledge the computing resources provided on Bebop, a high-performance computing cluster operated by the Laboratory Computing Resource Center at Argonne National Laboratory and NREL high-performance computing facility.


\section*{Graphical abstract}
\begin{figure}[htbp]
	\centering
	\includegraphics[width=10cm]{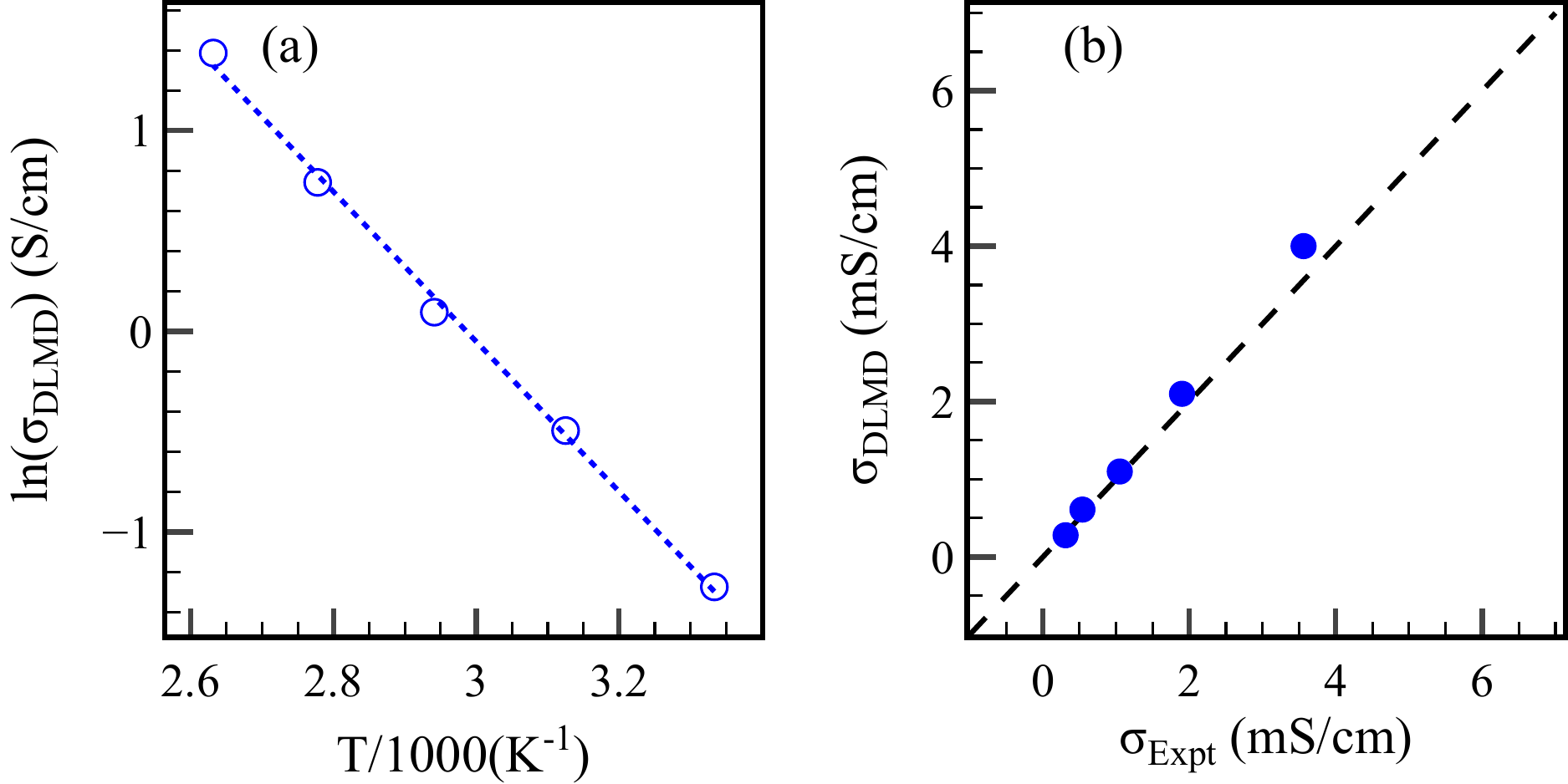}
	\vspace{1mm}
	\caption{Graphical abstract. (a) Ionic conductivity of LPS:Ti$_{20\%}$ (b) Calculated ionic conductivity versus experimental ionic conductivity with temperature from 25$^\mathrm{o}$C to 225$^\mathrm{o}$C.}
	\label{graphical_abstract}
\end{figure}	

\phantomsection %
\addcontentsline{toc}{part}{References}
\bibliography{ms}

\clearpage

\setcounter{section}{0}
\setcounter{figure}{0}  
\setcounter{table}{0}  
\renewcommand{\theHsection}{Supplement.\thesection} %
\renewcommand{\theHsubsection}{Supplement.\thesubsection}  
\renewcommand{\theHsubsubsection}{Supplement.\thesubsubsection}  
\renewcommand{\thefigure}{S\arabic{figure}}  
\renewcommand{\thetable}{S\arabic{table}}  
\renewcommand{\thesection}{S\arabic{section}}  
\renewcommand{\thesubsection}{S\arabic{section}.\arabic{subsection}}  
\renewcommand{\thesubsubsection}{S\arabic{section}.\arabic{subsection}.\arabic{subsubsection}}

\phantomsection %
\addcontentsline{toc}{part}{Supplementary Information}

	
	




\end{document}